\documentclass[conference]{IEEEtran}
\IEEEoverridecommandlockouts

\usepackage{cite}
\usepackage{amsmath,amssymb,amsfonts}
\usepackage{graphicx}
\usepackage{textcomp}
\usepackage{xcolor}
\def\BibTeX{{\rm B\kern-.05em{\sc i\kern-.025em b}\kern-.08em
    T\kern-.1667em\lower.7ex\hbox{E}\kern-.125emX}}

\usepackage{cleveref}
\usepackage{tabularray}
\usepackage{booktabs}
\usepackage{multirow}
\usepackage{makecell}

\usepackage{tabularray}
\usepackage{booktabs}
\usepackage{multirow}
\usepackage{makecell}

\usepackage[ruled,vlined]{algorithm2e}
\usepackage{algpseudocode}
\makeatletter
\def\BState{\State\hskip-\ALG@thistlm}
\makeatother
\newtheorem{definition}{Definition}

\usepackage{matlab-prettifier}
\usepackage{color}
\usepackage{mathrsfs}
\usepackage[shortlabels]{enumitem}
\usepackage{subcaption}

\newcommand{\Continue}{\textbf{continue}}

\begin{document}

%\title{Game-Theoretic Timing Strategies\\ for Cyber Deception
%}

%\author{\IEEEauthorblockN{Ya-Ting Yang}
%\IEEEauthorblockA{\textit{Department of Electrical and Computer Engineering} \\
%\textit{Tandon School of Engineering}\\
%\textit{New York University}\\
%Brooklyn, NY, USA \\
%yy4348@nyu.edu}
%\and
%\IEEEauthorblockN{Quanyan Zhu}
%\IEEEauthorblockA{\textit{Department of Electrical and Computer Engineering} \\
%\textit{Tandon School of Engineering}\\
%\textit{New York University}\\
%Brooklyn, NY, USA \\
%qz494@nyu.edu}
%}

\title{\LARGE \bf Internet of Agentic AI: Incentive-Compatible Distributed Teaming and Workflow
}

\author{
Ya-Ting Yang  and  Quanyan Zhu\\
Department of Electrical and Computer Engineering, 
Tandon School of Engineering, \\ New York University, 
Brooklyn, NY, USA; 
\texttt{\{yy4348, qz494\}@nyu.edu}
}

\maketitle

\begin{abstract}
Large language models (LLMs) have enabled a new class of agentic AI systems that reason, plan, and act by invoking external tools. However, most existing agentic architectures remain centralized and monolithic, limiting scalability, specialization, and interoperability. This paper proposes a framework for scalable agentic intelligence, termed the Internet of Agentic AI, in which autonomous, heterogeneous agents distributed across cloud and edge infrastructure dynamically form coalitions to execute task-driven workflows. We formalize a network-native model of agentic collaboration and introduce an incentive-compatible workflow-coalition feasibility framework that integrates capability coverage, network locality, and economic implementability. To enable scalable coordination, we formulate a minimum-effort coalition selection problem and propose a decentralized coalition formation algorithm. The proposed framework can operate as a coordination layer above the Model Context Protocol (MCP). A healthcare case study demonstrates how domain specialization, cloud-edge heterogeneity, and dynamic coalition formation enable scalable, resilient, and economically viable agentic workflows. This work lays the foundation for principled coordination and scalability in the emerging era of Internet of Agentic AI.
\end{abstract}

\begin{IEEEkeywords}
Agentic AI, workflow, incentive compatibility, coalition formation
\end{IEEEkeywords}

\section{Introduction}

The emergence of Large Language Models (LLMs), together with their increasing integration with external tools, APIs, and memory systems, has given rise to a new paradigm of agentic AI, systems that move beyond passive information processing to actively reason, plan, and act in pursuit of high-level objectives \cite{acharya2025agentic}. This paradigm is particularly effective when LLMs are embedded in modular architectures, where distinct components specialize in task planning, tool invocation, information retrieval, verification, and evaluation. Recent work has shown that such multi-agent orchestration enables problem-solving
capabilities that exceed those of a single monolithic model \cite{wu2024autogen,shu2024towards,tran2025multi,tian2025beyond}.

A critical enabler of these systems is the Model Context Protocol (MCP), which provides a standardized interface for agents to dynamically discover and invoke external tools. By abstracting tool schemas, declared capabilities, and invocation endpoints, MCP allows workflows to adapt at runtime rather than relying on static integrations. In principle, this design supports extensibility and reuse, enabling agents to leverage tools developed independently of the core model. Moreover, MCP accommodates heterogeneous ownership structures: tools may be proprietary, licensed, access-controlled, or implemented as autonomous agents deployed elsewhere on a network.

Despite interface-level flexibility, most agentic AI systems are deployed within centralized and bounded computational environments. While such design simplifies integration and enables low-latency coordination, it fundamentally limits scalability.
In particular, scalability is constrained by assumptions of centralized ownership and control, limited organizational capacity to develop and govern diverse expert agents, and poor interoperability across users, policies, and different domains. As a result, current systems scale primarily in model size or computation, rather than in functional diversity, organizational reach, or capability composition.
Unlocking the next stage of agentic AI requires a shift from centrally orchestrated deployments to networked agent ecosystems, where agents and tools can be discovered, coordinated, and composed across organizational boundaries. This re-frames scalability as a system-level challenge involving coordination, trust, access control, and incentives, rather than computation alone.

\begin{figure}
    \centering
    \includegraphics[width=0.9\linewidth]{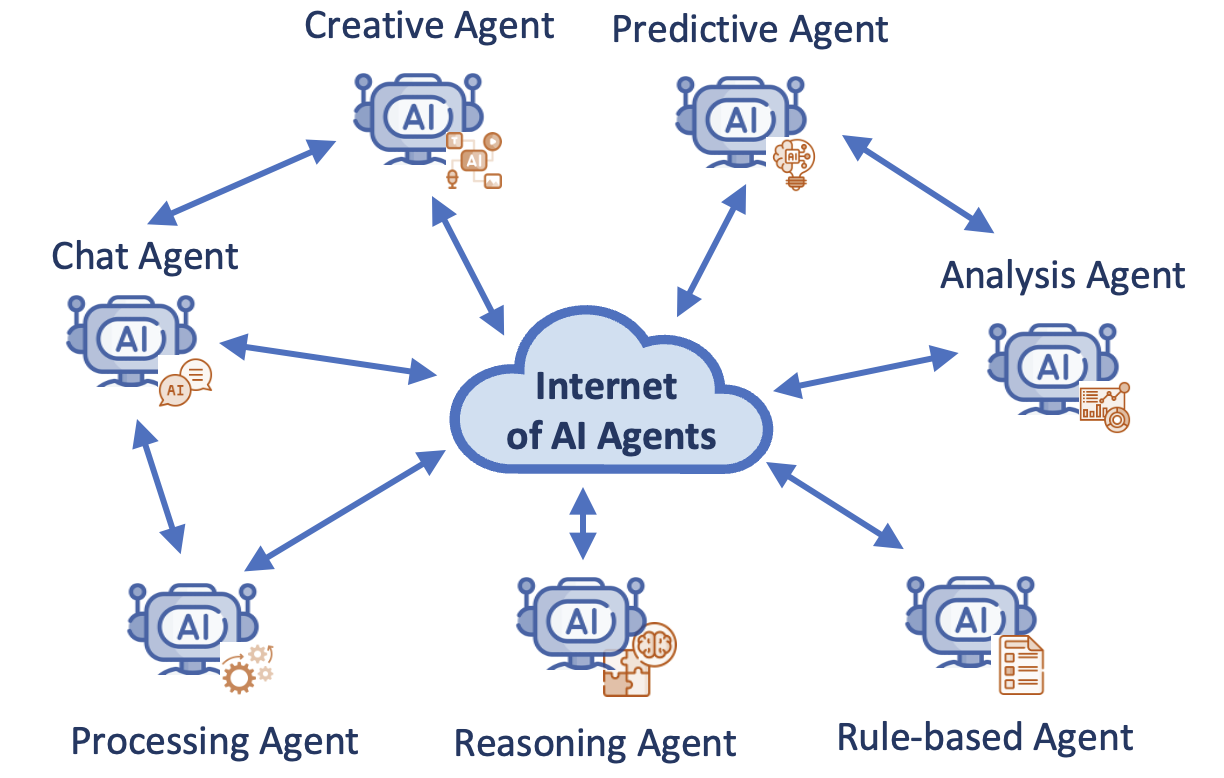}
    \caption{An illustration of the Internet of Agentic AI. Each node hosts one or more specialized AI agents designed for different tasks and participates in a shared communication network.}
    \label{fig:internet}
\vspace{-5mm}
\end{figure}

To overcome the scalability limitations of centralized agentic AI, we propose an architectural vision: the Internet of Agentic AI. In this framework, agentic intelligence is not confined to a single deployment or administrative domain, but instead emerges from a network of distributed nodes that may be deployed across cloud platforms, edge infrastructures, and institutional environments. Each node hosts one or more specialized AI agents and participates in a shared communication fabric, as illustrated in Fig. \ref{fig:internet}. A node may announce a task, such as investigating a cybersecurity incident or optimizing a supply chain into the network. Other nodes discover the task, evaluate its compatibility with their own capabilities, and dynamically form a coalition to collectively execute the task workflow. Upon task completion, rewards are allocated according to predefined mechanisms.

Our contribution can be summarized as follows. First, we propose a network-native model of agentic collaboration, where tasks arise dynamically and are specified by capability requirements. Agents discover over the network and form temporary coalitions spanning organizational and cloud–edge boundaries to execute the task-specific workflow.
Second, we introduce an incentive-compatible workflow–coalition feasibility framework that couples coalition formation with distributed workflow execution, characterizing when coalitions are feasible and workflows are executable under network constraints and budgets.
Third, we study the minimum-effort coalition selection problem and present a decentralized coalition-formation algorithm using only local network information, making it suitable for open ecosystems.
Finally, we discuss that the proposed framework can serve as a coordination layer that determines coalition composition, workflow assembly, and reward feasibility before execution, positioning this work as a scalable, incentive-aware extension of existing MCP infrastructure.

\section{Related Works}
Coalition formation in multi-agent systems \cite{sarkar2022survey} has deep roots in cooperative game theory \cite{branzei2008models}. The classical model of transferable utility (TU) games introduced by Shapley~\cite{shapley1953value} and further extended in survey treatments such as Chalkiadakis et al.~\cite{chalkiadakis2011computational} provides foundational tools for understanding how agents can coordinate to achieve collective utility and divide rewards. Concepts such as the Shapley value, the core, and coalition stability have been widely adopted in distributed systems~\cite{sandholm1999coalition}. However, identifying desirable coalition structures is computationally challenging, as the space of possible coalitions grows with the number of agents. To address this, a broad range of algorithmic approaches have been proposed to compute feasible and high-utility coalitions \cite{rahwan2007anytime}. Recent studies also address the challenges of coalition formation under practical constraints such as communication costs and resource constraints~\cite{rahwan2007anytime, saad2009coalitional}. Graph-constrained coalition games~\cite{michalak2013efficient} specifically incorporate topological limits into coalition feasibility, which is highly relevant in our setting. Our study differs from this line of research in that we focus on forming a task-specific coalition, rather than optimizing a global partition of all agents. Building on these ideas, we couple coalition valuation with workflow-dependent outcomes, enabling a tighter integration of network structure, incentives, and task viability.

Incentive compatibility in decentralized AI ecosystems has emerged as an active and rapidly growing area of research, particularly in the contexts of federated and distributed learning \cite{tu2022incentive,zhan2021survey}. As learning and decision-making processes increasingly rely on contributions from multiple autonomous participants, aligning individual incentives with system-level objectives has become a central challenge. A substantial body of work has therefore focused on designing mechanisms that encourage truthful participation, fair reward allocation, and sustained cooperation among self-interested agents. For instance, Ghosh and Hummel~\cite{ghosh2016incentivizing} study truthful mechanisms for agent task participation, Zhan et al.~\cite{zhan2020learning} analyze reward allocation schemes that account for communication costs and agent contributions, and Qi et al.~\cite{qi2025towards} utilize a blockchain-based framework for transparent agent registration, verifiable task allocation, and reputation tracking by smart contracts.

The emergence of agentic AI architectures, driven by the capabilities of Large Language Models (LLMs), has opened new paradigms for distributed reasoning and autonomous coordination. Early modular agent systems like AutoGPT\footnote{https://github.com/Significant-Gravitas/Auto-GPT} and BabyAGI\footnote{https://babyagi.org/} have demonstrated the power of chaining LLMs for planning and task execution. More recently, frameworks such as CAMEL~\cite{liu2023camel}, AgentVerse~\cite{ma2024agentverse}, and Multi-Agent Debate~\cite{du2023multiagent} formalize collaborative and competitive multi-agent behaviors using LLMs.
These works, however, largely assume centralized orchestration. In contrast, our architecture introduces a decentralized alternative, where agent capabilities are distributed across a network and coordination occurs via incentive-compatible coalition formation. This aligns with the vision of Internet-native agentic computing~\cite{openai2024agentic} and enables scalable, interoperable AI workflows.

\section{Coalition over Networks}

\begin{figure}
    \centering
    \includegraphics[width=\linewidth]{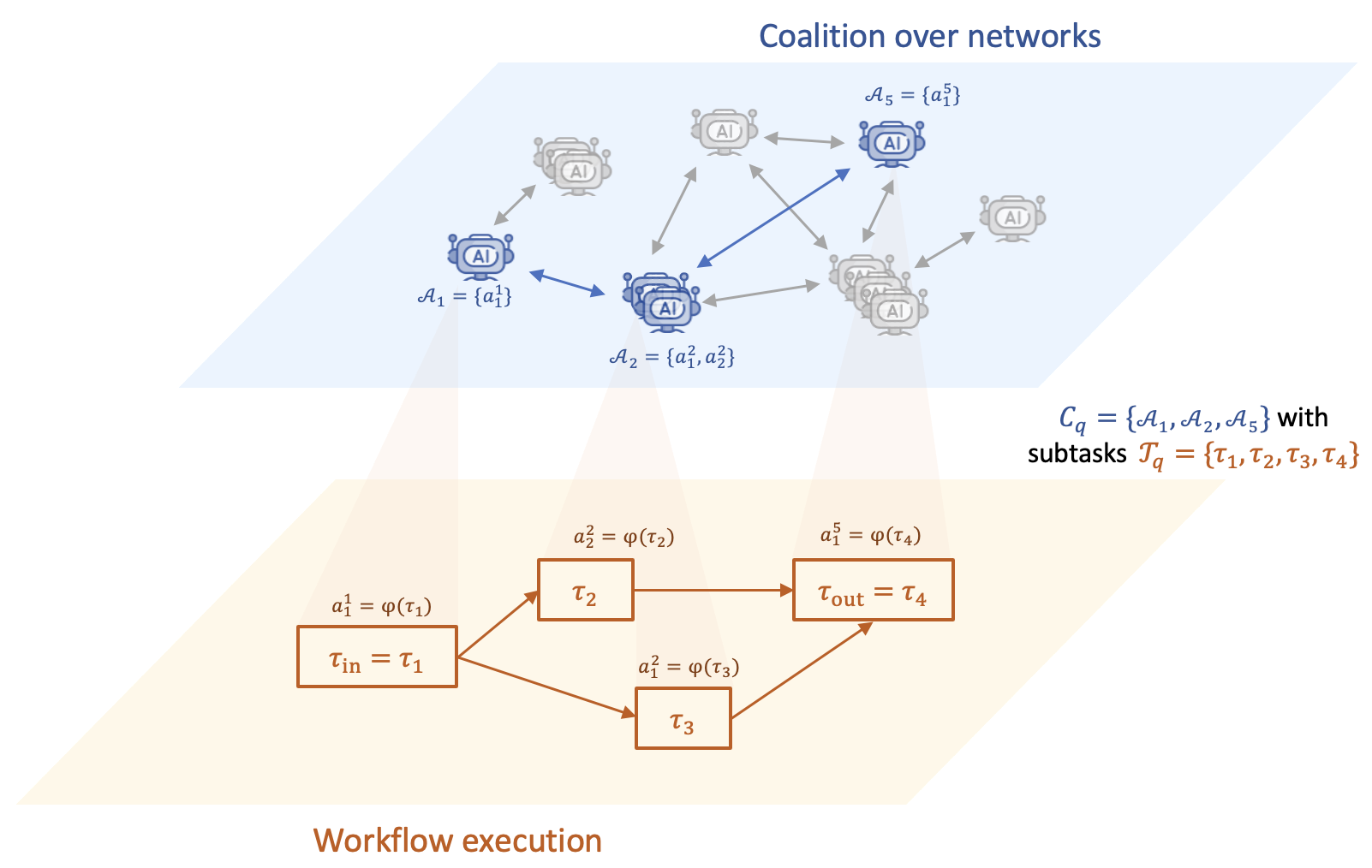}
    \caption{An illustration of task-specific capability-covering coalition formation with distributed workflow execution.}
    \label{fig:coalition-workflow}
\vspace{-5mm}
\end{figure}

We consider a distributed multi-agent system represented as a communication network, modeled by an undirected graph $\mathcal{G} = (\mathcal{V}, \mathcal{E})$,
where $\mathcal{V} = \{1, 2, \dots, N\}$ denotes the set of nodes and $\mathcal{E} \subseteq \mathcal{V} \times \mathcal{V}$ represents the set of communication edges between them.
Each node $i \in \mathcal{V}$ corresponds to an autonomous platform such as a computing unit, institution, or cyber-physical system that locally hosts a set of AI agents, denoted $\mathcal{A}_i = \{ a^i_1, a^i_2, \dots, a^i_{n_i} \}$. Each agent $a \in \mathcal{A}_i$ is endowed with a set of functional capabilities $C(a) \subseteq \mathcal{T}$, where $\mathcal{T} = \{c_1, c_2, \dots, c_M\}$ defines the global capability space, encompassing various competencies such as visual recognition, summarization, logical reasoning, or domain-specific inference.

\subsection{Domain Specialization and Task Requirements}

The system exhibits domain specialization. Each node $i$ is associated with a subset $\mathcal{T}_i \subseteq \mathcal{T}$ that characterizes its domain expertise. Agents hosted at node $i$ are assumed to draw their capabilities primarily from  $\mathcal{T}_i$, reflecting practical deployments where AI models are tailored to institutional contexts, for example, radiological reasoning agents in imaging centers or legal document interpreters at compliance hubs. This heterogeneity supports modularity and scalability, as nodes need only maintain models aligned with their specialization while relying on the network to fulfill other capabilities as needed.

Tasks are modeled as capability requirements and arise dynamically at nodes. A task $q$ initiated at node $i_0 \in \mathcal{V}$ is specified by a multiset of required capabilities: $\mathcal{R}_q = \{ c_1^{(r_1)}, c_2^{(r_2)}, \dots, c_M^{(r_M)} \}, r_k \in \mathbb{N}$,
indicating that $r_k$ number of agents with capability $c_k \in \mathcal{T}$ are needed to complete the task. If the initiating node $i_0$ is unable to satisfy $\mathcal{R}_q$ using its own agent pool $\mathcal{A}_{i_0}$, it launches a coalition formation process to recruit external support. The coalition is defined as a subset of nodes $\mathcal{C}_q \subseteq \mathcal{V}$, with the constraint $i_0 \in \mathcal{C}_q$, such that the collective agent set from nodes in $\mathcal{C}_q$ is sufficient to meet the capability requirements of the task $q$.

This formulation establishes the foundation for distributed capability discovery and cooperative task execution, where domain-specialized nodes dynamically form temporary coalitions to collectively complete tasks that exceed the scope of individual capability.

\begin{definition}[Capability-Covering Coalition]
Given a task $q$ initiated at node $i_0 \in \mathcal{V}$ with capability requirements $\mathcal{R}_q = \{ c_1^{(r_1)}, c_2^{(r_2)}, \dots, c_M^{(r_M)} \}, r_k \in \mathbb{N}$,
a subset of nodes $\mathcal{C}_q \subseteq \mathcal{V}$ is called a capability-covering coalition if:
i) $i_0 \in \mathcal{C}_q$ (i.e., the initiating node is included in the coalition), and ii) for each required capability $c_k \in \mathcal{T}$, the number of agents across all nodes in 
$\mathcal{C}_q$ possessing $c_k$ satisfies:
    $$
    \big| \big\{ a \in \bigcup_{j \in \mathcal{C}_q} \mathcal{A}_j \,\mid\, c_k \in C(a) \big\} \big| \geq r_k.
    $$
\label{def:capability-covering_coalition}
\vspace{-3mm}
\end{definition}

\begin{definition}[k-Degree Feasibility]
Given a communication network $\mathcal{G} = (\mathcal{V}, \mathcal{E})$, we say that the task $q$ with capability requirement multiset $\mathcal{R}_q$ initiated at node $i_0 \in \mathcal{V}$ is k-degree feasible if there exists a capability-covering coalition $\mathcal{C}_q \subseteq \mathcal{V}$ as defined in Definition \ref{def:capability-covering_coalition} such that every node $j \in \mathcal{C}_q$ satisfies $d(i_0, j) \leq k$, where $d(i_0, j)$ is the shortest-path distance in $\mathcal{G}$.
If no such coalition exists within $k$-hops of $i_0$, then the task $q$ is said to be $k$-degree infeasible.
\label{def:k-degree}
%\vspace{-3mm}
\end{definition}

\begin{definition}[Feasibility Radius]
Given a task $q$ initiated at node $i_0 \in \mathcal{V}$, the feasibility radius $\rho_q \in \mathbb{N} \cup \{\infty\}$ is defined as the smallest integer $k$ such that there exists a $k$-degree feasibility coalition $\mathcal{C}_q \subseteq \mathcal{V}$ as defined in Definition \ref{def:k-degree}.
If no such coalition exists within any finite distance, we set \( \rho_q = \infty \) and say that the task is globally infeasible on the network.
\end{definition}

\subsection{Distributed Workflow Execution}
Once a coalition $\mathcal{C}_q \subseteq \mathcal{V}$ is formed to complete a task $q$, the participating agents collaboratively execute a distributed workflow, as illustrated in Fig. \ref{fig:coalition-workflow}. The task $q$ is decomposed into a set of interdependent sub-tasks $\mathcal{T}_q = \{ \tau_1, \tau_2, \dots, \tau_L \}$, where each sub-task  $\tau_\ell \in \mathcal{T}_q$ requires a specific capability $c(\tau_\ell) \in \mathcal{T}$. The execution structure is modeled by a directed acyclic graph (DAG), denoted as $\mathcal{W}_q = (\mathcal{T}_q, \mathcal{D}_q)$, where the edge set $\mathcal{D}_q \subseteq \mathcal{T}_q \times \mathcal{T}_q$ encodes precedence relations: if $(\tau_i, \tau_j) \in \mathcal{D}_q$, then the output of $\tau_i$ serves as part of the input to $\tau_j$.
Each sub-task $\tau \in \mathcal{T}_q$ is assigned to an agent $a = \varphi(\tau) \in \bigcup_{j \in \mathcal{C}_q} \mathcal{A}_j$, such that the capability requirement is satisfied, i.e., $c(\tau) \in C(a)$. The assignment function is defined as $\varphi:\mathcal{T}_q \mapsto \bigcup_{j \in \mathcal{C}_q} \mathcal{A}_j$. Each agent $a \in \bigcup_{j \in \mathcal{C}_q} \mathcal{A}_j$ performs a local computation described by the tuple $(I_a, u_a, f_a, O_a)$, where $I_a \subseteq \mathcal{X}_{a}$ is the structured input from upstream agents, $u_a \in \mathcal{U}_a$ is a local control input (e.g., effort or configuration), and $f_a : \mathcal{X}_a \times \mathcal{U}_a \mapsto \mathcal{Y}_a$ is the local processing function. The output $O_a \in \mathcal{Y}_a$ is computed as $O_a = f_a(I_a, u_a)$.

\begin{definition}[Terminal Sub-Tasks]
A sub-task $\tau \in \mathcal{T}_q$ is said to be terminal if it has no outgoing edges, i.e., $\forall \tau' \in \mathcal{T}_q, (\tau, \tau') \notin \mathcal{D}_q$. The set of terminal sub-tasks is denoted
$$
\mathcal{T}_q^{\mathrm{term}} := \left\{ \tau \in \mathcal{T}_q \mid \nexists \tau' \in \mathcal{T}_q \text{ such that } (\tau, \tau') \in \mathcal{D}_q \right\}.
$$
\label{def:terminal_task}
\vspace{-3mm}
\end{definition}

\subsection{Compositional Workflow Dynamics}

The distributed workflow $\mathcal{W}_q = (\mathcal{T}_q, \mathcal{D}_q)$ induces a recursive composition of local agent computations. Each agent $a = \varphi(\tau)$ applies a function $f_a : \mathcal{X}_a \times \mathcal{U}_a \rightarrow \mathcal{Y}_a$, generating output $O_a = f_a(I_a, u_a)$, where $I_a = (O_{\varphi(\tau')})_{\tau' \in \mathsf{Pred}(\tau)}$ aggregates the outputs of predecessor tasks. The predecessor set is defined as $\mathsf{Pred}(\tau) := \{ \tau' \in \mathcal{T}_q \mid (\tau', \tau) \in \mathcal{D}_q \}$. This results in the recursive relation
$$
O_{\varphi(\tau)} = f_{\varphi(\tau)}\left( \left( f_{\varphi(\tau')}( \cdot , u_{\varphi(\tau')} ) \right)_{\tau' \in \mathsf{Pred}(\tau)}, u_{\varphi(\tau)} \right),
$$
which unfolds from source sub-tasks. For such sub-tasks $\tau$ where $\mathsf{Pred}(\tau) = \emptyset$, the input is an exogenous signal $x_\tau \in \mathcal{X}_{\varphi(\tau)}$, and we write $O_{\varphi(\tau)} = f_{\varphi(\tau)}(x_\tau, u_{\varphi(\tau)})$. The global task output is obtained by aggregating the terminal outputs:
$O_q = \mathsf{Agg} \left( \left\{ O_{\varphi(\tau)} \mid \tau \in \mathcal{T}_q^{\mathrm{term}} \right\} \right)$,
where $\mathsf{Agg}$ is a task-specific aggregation operator (e.g., concatenation, sum). This structure enables traceability, modularity, and principled performance analysis in distributed task execution.

\section{Incentive Compatibility and Coalition Games}

We model the collaboration of agentic AI teams across a networked system as a coalition-based game associated with the dynamically arising tasks.
Let $q \in \mathcal{Q}$ denote a task that appears at time $t$, and let $\mathcal{C}_q \subseteq \mathcal{V}$ be the capability-covering coalition (as defined in Definition \ref{def:capability-covering_coalition}) of nodes formed to complete task $q$.
Each node $i \in \mathcal{C}_q$ hosts a team of agents $\mathcal{A}_i = \{ a_{i1}, a_{i2}, \dots, a_{i n_i} \}$, and each agent $a \in \mathcal{A}_i$ contributes an effort $u_a^q \in \mathbb{R}_{\geq 0}$.
Then, the total effort from node $i$ is given by $u_i^q = \sum_{a \in \mathcal{A}_i} u_a^q$.
In addition, the collective (combined) agent effort profile across all participating nodes in $\mathcal{C}_q$ is denoted as $\mathbf{u}^q = \left( u_a^q \right)_{a \in \bigcup_{i \in \mathcal{C}_q} \mathcal{A}_i }$.
The combined agent effort then determines the task-specific outcome via a task outcome function $O_q = f_q(\mathbf{u}^q)$, and the resulting task reward is (designed as) $R_q = r_q(O_q) \in \mathbb{R}_{\geq 0}$.
Each node $i \in \mathcal{C}_q$ incurs i) effort cost $c_i(u_i^q)$, which is the aggregate cost of agent-level effort at node $i$, and ii) communication cost $C_i^{\text{comm}}=\sum_{j \in \mathcal{C}_q \setminus \{i\}} \gamma_{ij}^q$, where $\gamma_{ij}^q$ is the communication cost between nodes $i$ and $j$.
Let $w_i^q \in \mathbb{R}_{\geq 0}$ denote the reward allocated to node $i$, satisfying the budget-balanced constraint $\sum_{i \in \mathcal{C}_q} w_i^q = R_q$.
Then the net utility for node $i$ is
\begin{equation}
\pi_i^q = w_i^q - c_i(u_i^q) - C_i^{\text{comm}} .
\label{eq:node-utility-q}
\end{equation}

Based on the nodes' utilities, we say that a coalition $\mathcal{C}_q \subseteq \mathcal{V}$ is individually rational (IR) if each node receives non-negative utility:
\begin{equation}
\pi_i^q \geq 0 \quad \forall i \in \mathcal{C}_q .
\label{eq:ir-condition}
\end{equation}
It is incentive compatible (IC) if staying within the current coalition $\mathcal{C}_q$ yields at least as much utility as other alternative option:
\begin{equation}
\pi_i^q \geq \pi_i^{\text{out}} \quad \forall i \in \mathcal{C}_q ,
\label{eq:ic-condition}
\end{equation}
where \( \pi_i^{\text{out}} \) denotes the utility of abstaining from the current coalition or deviating to an alternative coalition. In the subsequent analysis, we treat $\pi_i^{\text{out}}=0, \forall i \in \mathcal{C}_q$ under the assumption that task $q$ is the only task at time $t$; hence, not participating in $\mathcal{C}_q$ yields zero utility. In this case, the IC constraints are equivalent to the IR constraints. We will leave the scenario with multiple tasks arising simultaneously (leading to multiple coalitions and $\pi_i^{\text{out}}\neq 0$) for future work.

\subsection{Consistency Between Coalition and Workflow}

One can observe that there is a fundamental coupling between coalition formation, workflow execution, and reward design. The outcome of the workflow determines the total task reward, which in turn governs the feasibility (individual rationality and incentive compatibility) of the coalition.
Conversely, the structure of the coalition, e.g., the participating agents and their allocated efforts, determines whether the workflow can be successfully executed. We say that the system is workflow-coalition feasible when this mutual dependence is satisfied: the workflow is executable given the coalition structure, and the resulting reward enables an implementable incentive scheme for all coalition members. 

\begin{definition}[Workflow-Coalition Feasibility]
\label{def:workflow-coalition-feasibility}
Let $q$ be a task initiated at node $i_0 \in \mathcal{V}$, capability requirements $\mathcal{R}_q$, and reward function $r_q$. The associated workflow  $\mathcal{W}_q = (\mathcal{T}_q, \mathcal{D}_q)$ is composed of interdependent sub-tasks $\tau \in \mathcal{T}_q$, each requiring an effort $u_a$ from an agent $a = \varphi(\tau)$ assigned to that sub-task. Let  $\mathcal{C}_q \subseteq \mathcal{V}$ be a proposed coalition of nodes, and let $\mathcal{A}_q = \bigcup_{i \in \mathcal{C}_q} \mathcal{A}_i$ denote the set of agents available for task $q$. Then, the pair $(\mathcal{W}_q, \mathcal{C}_q)$ is said to be workflow-coalition feasible if the following conditions are satisfied:
\begin{enumerate}
    \item[(ii)] \textbf{Capability-Covering:}
    $\mathcal{C}_q$ is a capability-covering coalition as defined in Definition \ref{def:capability-covering_coalition}.
    
    \item[(ii)] \textbf{Capability-Consistent:}
    There exists an assignment $\varphi: \mathcal{T}_q \mapsto \mathcal{A}_q$ such that for every $\tau \in \mathcal{T}_q$, the assigned agent $a = \varphi(\tau)$ satisfies the required capability: $c(\tau) \in C(a)$.

    \item[(iii)] \textbf{Well-Defined Workflow Output:}
    The recursive application of sub-task functions
    \( f_{\varphi(\tau)} \), driven by efforts \( \{ u_a \}_{a \in \mathcal{A}_q} \), yields a well-defined global outcome \( O_q \in \mathbb{R} \). That is, $O_{\varphi(\tau)}=f_{\varphi(\tau)}\!\left(I_{\varphi(\tau)}, u_{\varphi(\tau)}\right)$, with the terminal outputs aggregated as $O_q=\mathrm{Agg}\!\big(\{ O_{\varphi(\tau)} \}_{\tau \in \mathcal{T}_q^{\mathrm{term}}}\big)$.

    \item[(iv)] \textbf{Reward Realizability:}
    The task reward induced by the workflow outcome, $R_q = r_q(O_q)$, is finite and nonnegative: $0 \le R_q < \infty$.

    \item[(v)] \textbf{Budget Feasibility:}
    The total operational and communication costs incurred by coalition members can be covered by the task reward: $\sum_{i \in C_q}\bigl[ c_i(u_i) + C_i^{\mathrm{comm}} \bigr]\le R_q$.

    \item[(vi)] \textbf{Incentive Compatibility:}
    There exists a reward allocation
    $\{ w_i^q \}_{i \in \mathcal{C}_q}$ satisfying $w_i^q \ge c_i(u_i) + C_i^{\mathrm{comm}}, \forall i \in \mathcal{C}_q$, with $\sum_{i \in \mathcal{C}_q} w_i^q = R_q$.
\end{enumerate}
\end{definition}

\begin{definition}[$k$-Degree Workflow-Coalition Feasibility]
\label{def:k-degree-feasibility}
Let $G = (\mathcal{V}, \mathcal{E})$ denote the communication graph, and let $d(i,j)$ denote the shortest-path distance between nodes $i$ and $j$ in $G$.
The task $q$ is said to be $k$-degree workflow-coalition feasible if there exists a workflow-coalition feasible pair $(\mathcal{W}_q, \mathcal{C}_q)$ as defined in Definition \ref{def:workflow-coalition-feasibility} such that $d(i_0, j) \le k, \forall j \in \mathcal{C}_q$.
If no such coalition exists for a given $k$, then task $q$ is said to be $k$-degree infeasible. If no such $k \in \mathbb{N}$ exists, then task $q$ is said to be globally infeasible in the Internet of Agentic AI.

\end{definition}

\subsection{Minimum-Effort Coalition}
\label{subsec:min-effort-coalition}

In this section, we aim to identify a coalition that supports the given workflow while minimizing the total effort exerted by participating agents.

\begin{definition}[Minimum-Effort Coalition Selection]
\label{def:min-effort-coalition}
Let $q$ be a task with associated workflow $\mathcal{W}_q = (\mathcal{T}_q, \mathcal{D}_q)$, initiated at node $i_0 \in \mathcal{V}$. For a given hop radius $k$, define
\begin{equation}
\mathcal{C}_q^{(k)} \triangleq \Bigl\{\mathcal{C}_q \subseteq \mathcal{V}\mid \mathcal{C}_q \text{ satisfies Definition \ref{def:k-degree-feasibility}}\Bigr\}.
\label{eq:kdegree-coalitions}
\end{equation}
The Minimum-Effort Coalition Selection Problem seeks a coalition $\mathcal{C}_q^\star \in \mathcal{C}_q^{(k)}$ that minimizes the total agentic effort required to execute the workflow:
\begin{equation}
\mathcal{C}_q^\star \in \arg\min_{\mathcal{C}_q \in \mathcal{C}_q^{(k)}} \sum_{i \in \mathcal{C}_q} \sum_{a \in \mathcal{A}_i} u_a.
\label{eq:min-effort-objective}
\end{equation}
If $\mathcal{C}_q^{(k)} = \emptyset$, no feasible coalition exists at hop radius $k$. In this case, the system may increase $k$ iteratively until a feasible coalition is identified or a predefined exploration limit is reached.
\end{definition}

\begin{algorithm}
\caption{Workflow-Coalition Formation}
\label{alg:workflow-coalition}
\KwIn{Task $q$ initiated at node $i_0$, workflow $\mathcal{W}_q = (\mathcal{T}_q, \mathcal{D}_q)$,
capability requirements $\mathcal{R}_q$, $\{ c(\tau_\ell) \}_{\ell=1}^L$,
maximum search radius $k_{\max}$}
\KwOut{Feasible coalition $\mathcal{C}_q^\star$ and workflow assignment, or \texttt{Infeasible}}

Initialize $k\leftarrow 1$\;

\While{ $k \le k_{\max}$ }{

    %\tcp{Step 1: Neighborhood exploration}
    Compute the $k$-hop neighborhood of node $i_0$:
    %\vspace{-5mm}
    $\mathcal{N}^{(k)}(i_0):=\{ i \in \mathcal{V} \mid d(i,i_0) \le k \}$.
    
    %\tcp{Step 2: Candidate coalitions}
    Construct the set of candidate coalitions for $\mathcal{R}_q$:
    %\vspace{-5mm}
    $\mathcal{F}_q^{(k)}:= \bigl\{\mathcal{C}_q \subseteq \mathcal{V}\mid \mathcal{C}_q \text{ satisfies Definition \ref{def:k-degree}}\bigr\}$.

    \If{\( \mathcal{F}_q^{(k)} = \emptyset \)}{
        \( k \leftarrow k+1 \); \Continue
    }

    %\tcp{Step 3: Workflow feasibility}
    Initialize \( \mathcal{C}_q^{(k)} \leftarrow \emptyset \)\;

    \ForEach{\( \mathcal{C}_q \in \mathcal{F}_q^{(k)} \)}{

        \If{$(\mathcal{W}_q, \mathcal{C}_q)$ is workflow-coalition feasible
        (Definition~\ref{def:workflow-coalition-feasibility}) with an assignment $\varphi$}{
            Add $\mathcal{C}_q$ to $\mathcal{C}_q^{(k)}$\;
        }
    }

    \If{ $\mathcal{C}_q^{(k)} = \emptyset$}{
        $k \leftarrow k+1$; \Continue
    }

    %\tcp{Step 4: Minimum-effort selection}
    Select $\mathcal{C}_q^\star$ as in Definition~\ref{def:min-effort-coalition}:
    %\vspace{-5mm}
    $\mathcal{C}_q^\star
    \in
    \arg\min_{\mathcal{C}_q \in \mathcal{C}_q^{(k)}}
    \sum_{i \in \mathcal{C}_q}
    \sum_{a \in \mathcal{A}_i} u_a$.
    
    \Return $(\mathcal{C}_q^\star, \mathcal{W}_q)$\;
}
\Return \texttt{Infeasible}
\end{algorithm}

The Workflow-Coalition Formation Algorithm, as described in Algorithm \ref{alg:workflow-coalition}, proposes a decentralized search-and-select process through which an agent initiates a task at a node $i_0$, incrementally discovers and evaluates candidate coalitions within expanding hop-radius neighborhoods.
At each radius $k$, the algorithm enumerates coalitions $\mathcal{C}_q \subseteq \mathcal{N}^{(k)}(i_0)$ that satisfy the locality constraint of $k$-degree feasibility (Definition~\ref{def:k-degree}). For each candidate coalition, the algorithm attempts to assign workflow sub-tasks to coalition members in a capability-consistent manner and verifies whether the resulting assignment satisfies workflow-coalition feasibility (Definition~\ref{def:workflow-coalition-feasibility}), including reward realizability, budget feasibility, and incentive compatibility.

If at least one feasible coalition is identified, the algorithm selects the coalition that minimizes total agent effort, thereby solving the Minimum-Effort Coalition Selection Problem (Definition~\ref{def:min-effort-coalition}). If no feasible coalition exists at the current hop radius, the search radius is expanded to $k+1$ until a feasible solution is found or a predefined maximum exploration radius $k_{\max}$ is reached. This procedure tightly couples workflow planning with decentralized coalition formation, ensuring that emergent tasks are supported by coalitions that are both structurally viable and economically implementable in the Internet of Agentic AI.

\section{Case Study: Internet of Agentic AI in Healthcare}

A compelling application of the Internet of Agentic AI arises in collaborative healthcare \cite{milne2020effectiveness}. In a regional medical ecosystem, institutions such as primary care clinics, diagnostic imaging centers, specialist hospitals, telehealth providers, and insurance platforms operate heterogeneous AI capabilities aligned with their institutional roles. These capabilities are deployed across a mix of edge and cloud infrastructure: clinics and telehealth providers often host lightweight, latency-sensitive agents at the edge, while imaging centers, hospitals, and insurers may operate compute-intensive LLM-based agents in the cloud, usually fine-tuned on domain-specific data such as radiology reports, pharmaceutical records, and clinical guidelines.

No single institution possesses the full spectrum of capabilities required to handle complex, multi-stage patient cases. Differences in expertise, data availability, privacy constraints, and computational resources make centralized solutions impractical. The Internet of Agentic AI enables these institutions to form dynamic, task-driven coalitions that combine complementary capabilities across the network.

\begin{figure}
    \centering
    \includegraphics[width=\linewidth]{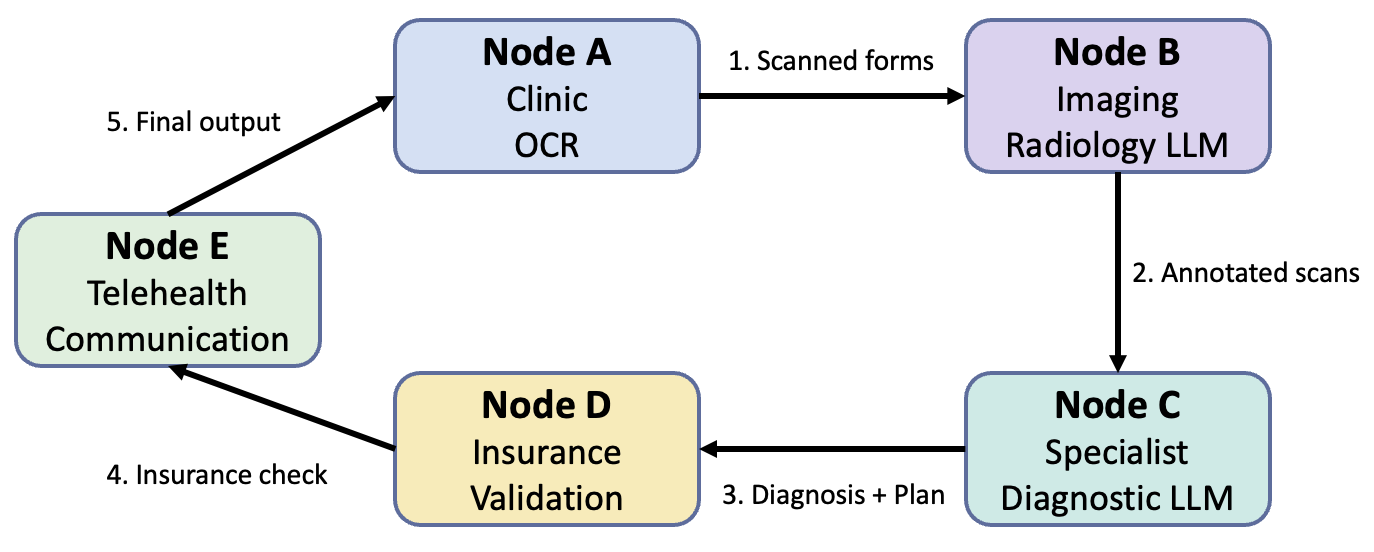}
    \caption{Agentic healthcare workflow with directional information flow. The system forms a closed loop in which the downstream diagnostic, validation, and communication outputs are returned to the clinic for coordination and finalization.}
    \label{fig:agentic-healthcare-workflow}
\vspace{-5mm}
\end{figure}

\subsection{Healthcare Workflow}
Figure \ref{fig:agentic-healthcare-workflow} shows a representative distributed agentic healthcare workflow for illustrative purposes. A patient presents at a small primary care clinic, denoted as Node~A, providing handwritten intake forms, radiology images, and prior lab results, along with ambiguous symptoms requiring specialized evaluation. Node~A hosts agents capable of optical character recognition (OCR) and structured form parsing, but often lacks advanced diagnostic capabilities.

Recognizing these limitations, Node~A initiates a distributed workflow by forming a coalition with other nodes. As shown in Step~1 of Fig. \ref{fig:agentic-healthcare-workflow}, OCR agents at Node~A digitize intake forms and extract structured information, which is then forwarded to Node~B, a regional imaging center hosting cloud-based radiology agents. These LLM-driven agents analyze medical images, generate annotated scans, and identify abnormalities with precise spatial localization (Step~2).

The workflow proceeds to Node~C, a specialist hospital whose cloud-deployed diagnostic LLMs perform differential diagnosis, rank candidate conditions, and recommend additional tests when needed (Step~3). Leveraging domain-specific clinical protocols, these agents provide context-aware assessments that exceed the reliability of general-purpose models.

Once a provisional diagnosis and treatment plan are synthesized, Node~A consults Node~D, an insurance and pharmacy integration platform. Agents at Node~D verify coverage, screen for drug interactions, and confirm medication availability (Step~4). Finally, Node~E, a telehealth provider, completes the workflow by generating patient-facing explanations, consent documentation, and scheduling follow-up consultations, returning the finalized output to Node~A for coordination and delivery (Step~5).

\subsection{Numerical Setups}\label{subsec:setup}

According to the workflow in Fig. \ref{fig:agentic-healthcare-workflow}, the five sub-tasks are mapped to five institutions:
\[
\begin{aligned}
\tau_1 &: \text{OCR and intake at Clinic } (A), \\
\tau_2 &: \text{Radiology interpretation (RAD) at Imaging Center } (B), \\
\tau_3 &: \text{Differential diagnosis (DX) at Specialist Hospital } (C), \\
\tau_4 &: \text{Policy validation (VAL) at Insurance Provider } (D), \\
\tau_5 &: \text{Teleconsultation (CONS) at Telehealth } (E).
\end{aligned}
\]
Let $u_i = u_\ell$ and $i = i_\ell$ for each $\ell$, due to the one-to-one mapping of agents to nodes. Each node incurs an operational cost $c_i(u_i)$ and a communication overhead $C_i^{\text{comm}} \geq 0$. The platform must allocate reward shares  $\{w_i^q\}_{i \in \mathcal{V}}$ to satisfy both incentive and budgetary constraints.

\subsubsection{Agentic effort (normalized units)}
For each node \( i \in \mathcal V \), let
$u_i \in \mathbb{R}_{\ge 0}$
denote the agentic effort expended at node $i$.
The variable $u_i$ is measured in normalized agentic effort units, where one unit corresponds to a fixed bundle of agentic computation and interaction, such as a reasoning--execution cycle, a tool-call round, or a standardized number of reasoning tokens.
More specifically, $u_i$ aggregates deliberation depth, interaction frequency, planning complexity, and inference load of the agents hosted at node $i$. This normalization allows heterogeneous agentic activities to be represented on a common scale while preserving the underlying cost-quality tradeoffs. Formally, $[u_i] = \text{normalized agentic effort units}$,
with physical quantities such as tokens, API calls, or GPU-seconds recoverable via a fixed scaling factor if desired.

\subsubsection{Effectiveness of agentic deliberation}

Let $g_i : \mathbb{R}_{\ge 0} \mapsto [0,1]$ denote the effectiveness of agentic deliberation at node $i$, mapping effort to a normalized measure of task-stage reliability.
We assume $g_i$ is increasing, continuously differentiable, and concave, capturing diminishing returns from additional deliberation. In the numerical studies, we adopt
\begin{equation}
g_i(u_i)=1 - e^{-\rho_i u_i},\ \rho_i > 0,
\label{eq:gi-agentic}
\end{equation}
where $\rho_i$ quantifies deliberation efficiency, i.e., the marginal improvement in reliability per unit of agentic effort.

\subsubsection{Workflow output}
Given the chain-structured healthcare workflow, the end-to-end task output is defined as
\begin{equation}
O_q(u)=\prod_{i \in \mathcal V} \alpha_i \, g_i(u_i),
\label{eq:Oq-agentic}
\end{equation}
where $\alpha_i \in (0,1]$ denotes the baseline reliability of the AI system or institutional process at node $i$.
The multiplicative structure reflects the compounding of errors across sequential stages of the healthcare pipeline, such as OCR, radiology, diagnosis, validation, and consultation.

\subsubsection{Reward function}
The task reward is given by
\begin{equation}
R_q(u)=\beta \log\!\left( 1 + O_q(u) \right),
\qquad \beta > 0,
\label{eq:Rq-agentic}
\end{equation}
where $R_q$ is measured in normalized utility units representing risk-adjusted healthcare value (e.g., reimbursement, avoided harm, or liability reduction).
The logarithmic form captures diminishing marginal value of increased decision confidence and reflects practical reimbursement ceilings.

\subsubsection{Cost of agentic execution}
Each node $i \in \mathcal V$ incurs a cost $c_i : \mathbb{R}_{\ge 0} \to \mathbb{R}_{\ge 0}$, modeled as
\begin{equation}
c_i(u_i)=\kappa_i^{\mathrm{cpu}} \, u_i+\kappa_i^{\mathrm{lat}} \, u_i^2,
\label{eq:ci-agentic}
\end{equation}
where $\kappa_i^{\mathrm{cpu}}$ (utility per unit effort) represents direct compute cost, such as GPU usage or API pricing; $\kappa_i^{\mathrm{lat}}$ (utility per unit effort squared) captures latency amplification, synchronization, and orchestration overhead that grows superlinearly with deliberation depth.
Each node also incurs a fixed communication cost $C_i^{\mathrm{comm}} \ge 0$, representing inter-agent messaging, secure data exchange, and institutional coordination overhead.

\subsection{Workflow–Coalition Formation}

\begin{figure}
    \centering
    \includegraphics[width=0.45\textwidth]{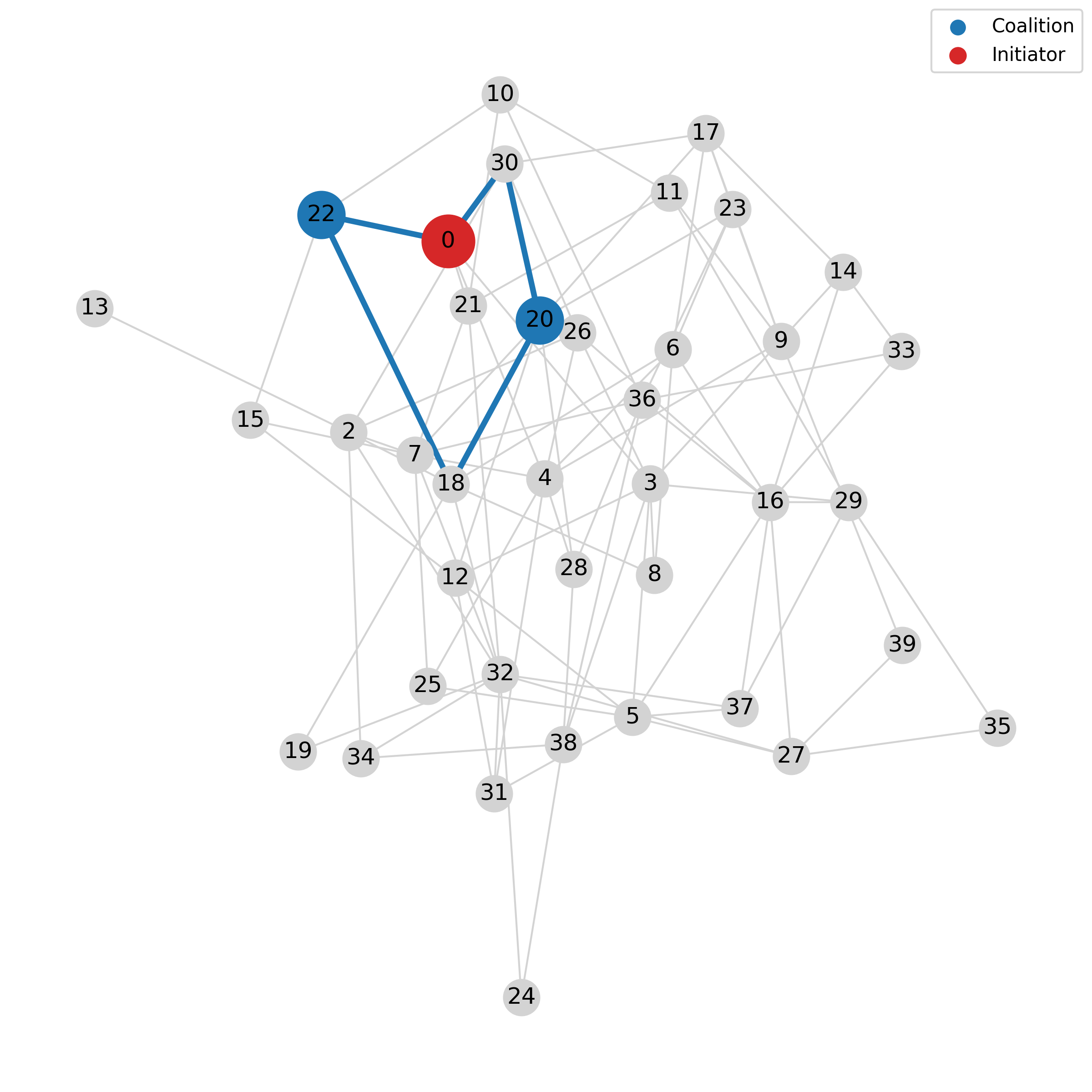}
    \caption{
    Example feasible workflow coalition.
    The initiator node is shown in red, and selected coalition members in blue.
    Highlighted edges indicate coalition-induced communication paths used to coordinate workflow execution.
    }
    \label{fig:network-coalition}
\vspace{-5mm}
\end{figure}

We consider an undirected Erd\H{o}s--R\'enyi graph with \(N=40\) nodes, where each node represents an agentic AI service endowed with a random subset of healthcare-relevant capabilities drawn from \(\{\mathrm{OCR}, \mathrm{RAD}, \mathrm{DX}, \mathrm{VAL}, \mathrm{CONS}\}\).
Each node is further parameterized by heterogeneous agentic characteristics, including deliberation efficiency, linear compute cost, quadratic latency penalty, fixed coordination overhead, and baseline effort level, as specified by the agentic cost and quality models in Section~\ref{subsec:setup}.

A workflow requiring all five capabilities is initiated at node $i_0 = 0$.
The coalition search-and-select algorithm (Algorithm \ref{alg:workflow-coalition}) explores neighborhoods of increasing hop radius up to a maximum of $k_{\max} = 4$, enumerating candidate coalitions that contain the initiating node, satisfy capability coverage, and meet the workflow-coalition economic feasibility condition (such as budget/reward feasibility and reduced IC/IR).

In the run visualized in Fig. \ref{fig:network-coalition}, the algorithm  identifies a feasible coalition at hop radius \(k = 2\). The selected coalition consists of nodes \(\{0, 20, 22\}\).
Based on the realized capability assignments, node \(0\) provides \(\{\mathrm{CONS}, \mathrm{OCR}\}\), node \(20\) provides \(\{\mathrm{DX}, \mathrm{RAD}\}\), and node \(22\) provides \(\{\mathrm{VAL}\}\), jointly covering all required workflow capabilities.
Among all feasible coalitions encountered during the search, this three-node coalition attains the minimum total agentic execution cost.
The figure also illustrates that full workflow feasibility can be achieved by a small, well-positioned coalition localized within a modest hop radius, despite the presence of many alternative nodes with overlapping capabilities in the network.

\begin{figure}[h!]
    \centering

    % -------------------------
    % (a) k vs x
    % -------------------------
    \begin{subfigure}[t]{0.48\textwidth}
        \centering
        \includegraphics[width=0.9\textwidth]{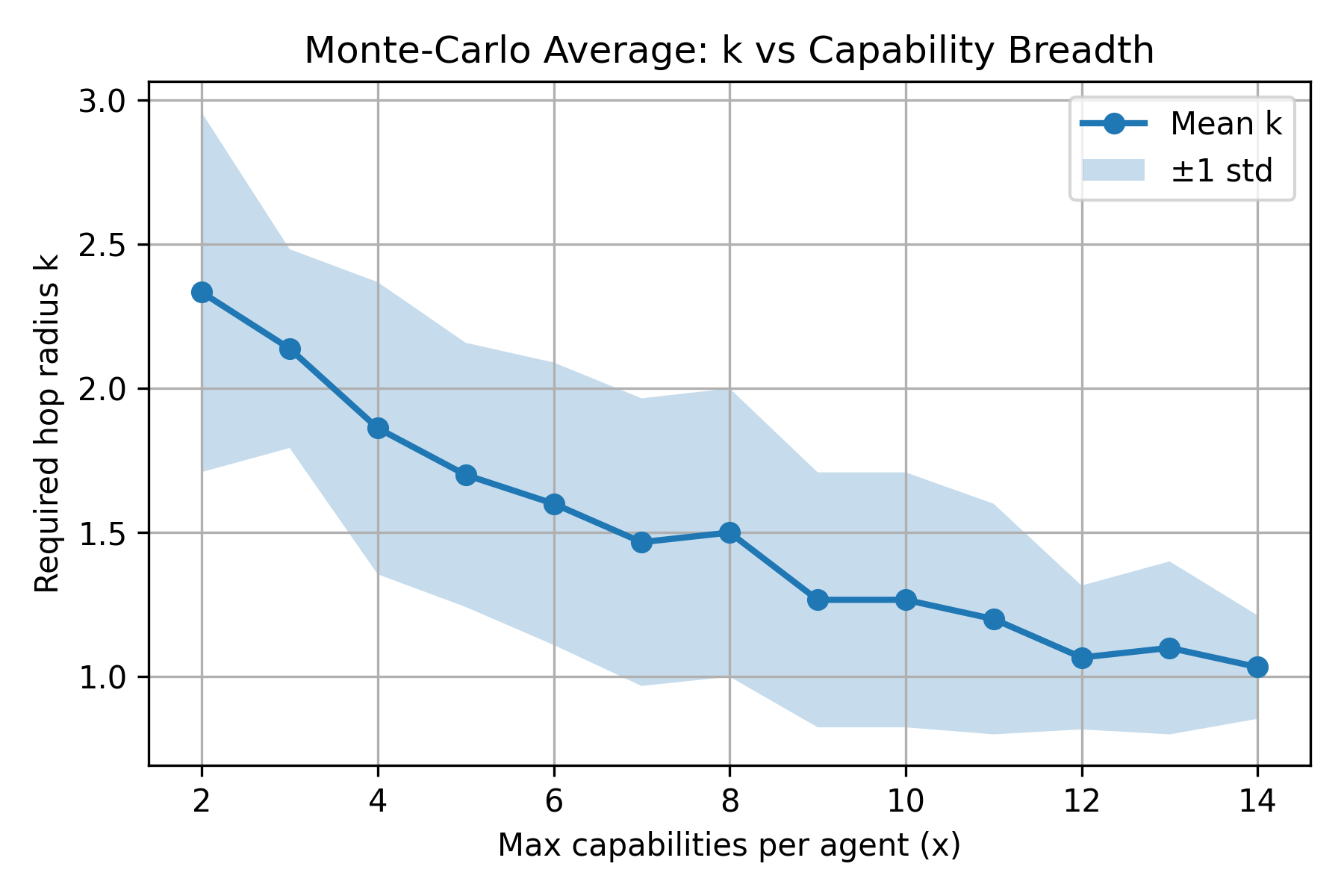}
        \caption{
        Average required hop radius \(k\) versus per-agent capability breadth \(x\). Shaded region denotes \(\pm 1\) standard deviation.
        }
        \label{fig:k-vs-x-mc}
    \end{subfigure}
    \hfill
    % -------------------------
    % (b) Coalition size vs x
    % -------------------------
    \begin{subfigure}[t]{0.48\textwidth}
        \centering
        \includegraphics[width=0.9\textwidth]{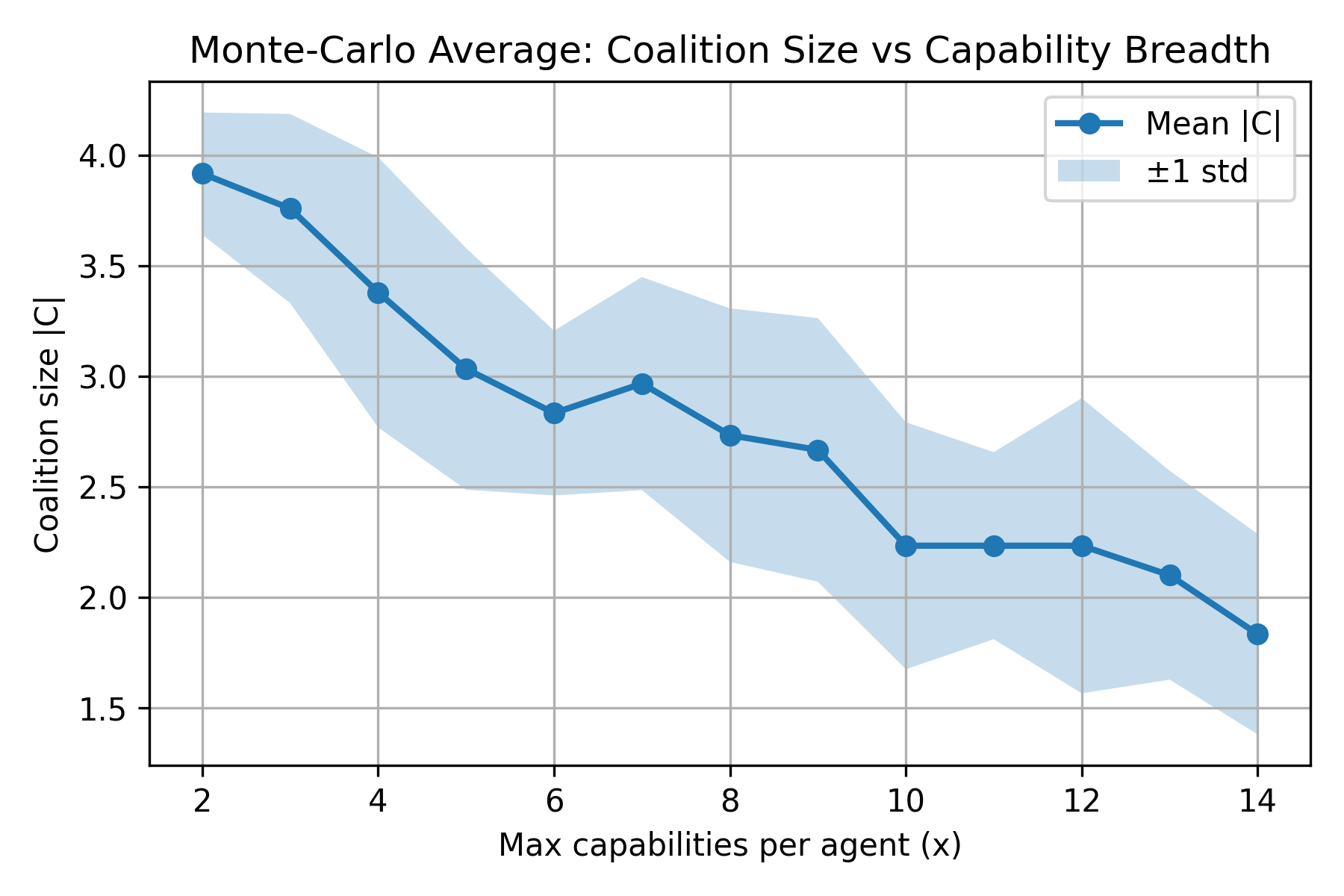}
        \caption{
        Average coalition size \(|C|\) versus capability breadth \(x\), illustrating reduced coordination requirements as agentic richness increases.
        }
        \label{fig:coalition-size-vs-x-mc}
    \end{subfigure}
    \hfill

    % -------------------------
    % (c) Cost vs iterations (single run only)
    % -------------------------
    \begin{subfigure}[t]{0.48\textwidth}
        \centering
        \includegraphics[width=0.9\textwidth]{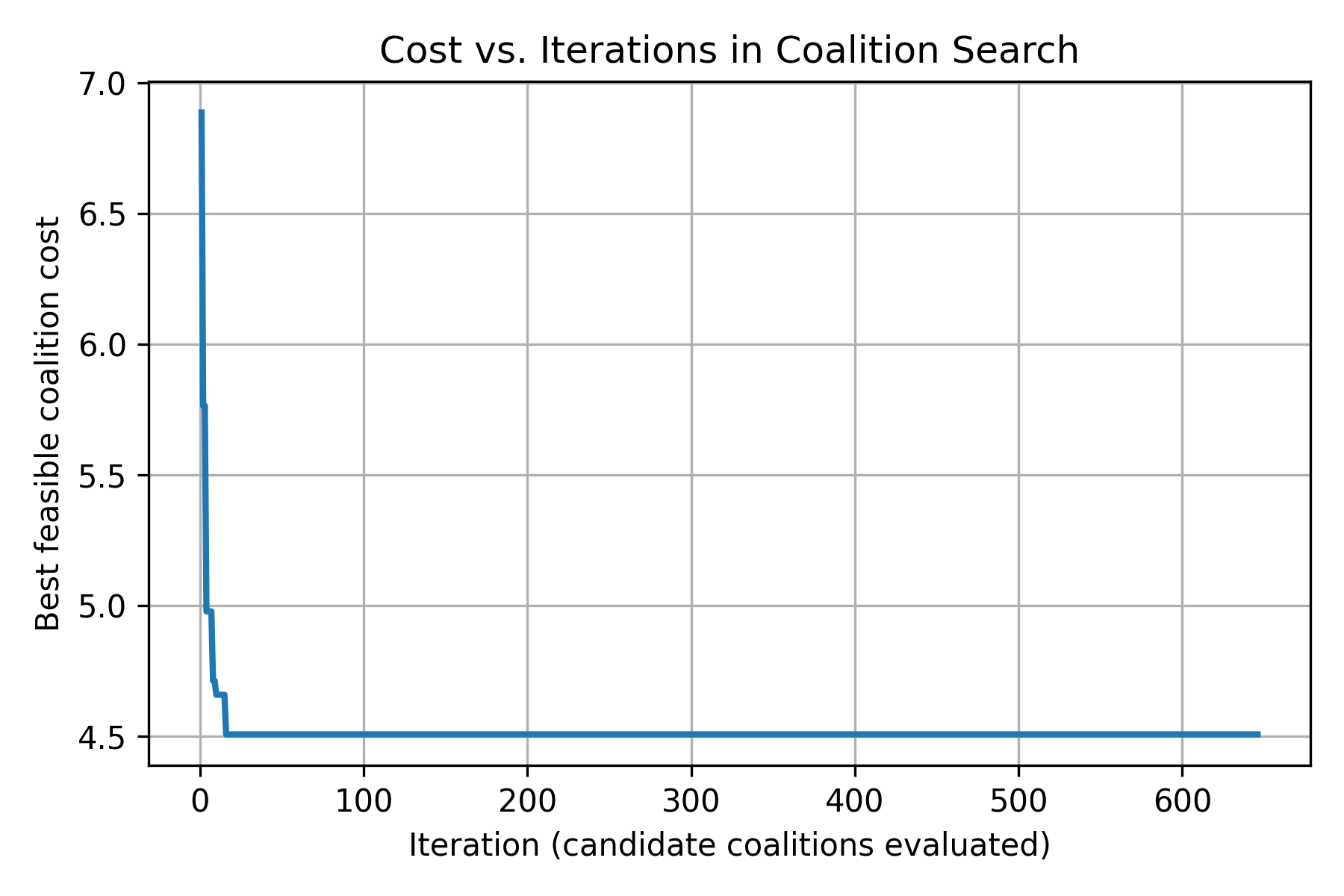}
        \caption{
        Best feasible coalition cost versus search iterations for a representative
        coalition formation run.
        }
        \label{fig:cost-vs-iter-single}
    \end{subfigure}

    \caption{
    Scaling behavior of workflow-coalition formation.
    Increasing per-agent capability breadth reduces both the coordination radius and coalition size required for feasibility.
    The coalition search exhibits rapid convergence in terms of the best achievable cost.
    }
    \label{fig:coalition-scaling}
\vspace{-5mm}
\end{figure}

The economic efficiency of the selected coalition is quantified by a total cost of \(4.51\) and a corresponding workflow reward of \(8.25\), yielding a strictly positive surplus.
The evolution of the best feasible coalition cost over the search process is shown in Figure~\ref{fig:cost-vs-iter-single}.
The curve exhibits a sharp initial decrease followed by stabilization, indicating that a near-optimal feasible coalition is discovered early in the search and that subsequent evaluations primarily confirm optimality rather than produce further improvements.

Beyond this single realization, the impact of agentic capability breadth on coordination requirements is examined via Monte-Carlo averaging. Figure~\ref{fig:k-vs-x-mc} reports the average hop radius required to achieve workflow feasibility as a function of the maximum number of capabilities per agent. As capability breadth increases, the required coordination radius decreases, revealing a substitution effect between local agentic richness and network-level coordination.
Similarly, Figure~\ref{fig:coalition-size-vs-x-mc} shows that the average coalition size decreases with capability breadth, confirming the intuition that richer agents reduce the need for coalition growth.

\section{Discussions: as an Extension of MCP}
\label{subsec:mcp-coalition}

The Model Context Protocol (MCP) provides a standardized, safety-oriented interface for discovering and invoking external tools by exposing their schemas, declared capabilities, and invocation endpoints. 
The workflow-coalition algorithm developed in this paper is compatible with MCP and does not modify its execution semantics. Instead, it operates strictly above MCP as a pre-execution coordination and planning layer.
We refer to this augmented architecture as
\(
\mathrm{C{+}MCP},
\)
where MCP remains the execution substrate, and \(\mathrm{C}\) denotes a coordination layer responsible for coalition formation, feasibility evaluation, and cost-aware tool selection.

Within the \(\mathrm{C^{+}MCP}\) interpretation, each MCP tool is modeled as an agentic entity that advertises not only its functional capabilities, but also abstract descriptors of effort-dependent quality, compute cost, latency, and coordination overhead. The initiating LLM acts as a leader node $i_0$, while MCP tools populate the surrounding network as candidate coalition members. These abstractions enrich the decision logic of tool selection without imposing any changes on MCP’s interface, schemas, or runtime behavior.

Specifically, coordination in \(\mathrm{C^{+}MCP}\) occurs prior to tool invocation. Given a workflow specification, the leader agent first solves a workflow-coalition feasibility problem: it searches over a bounded neighborhood of available tools to identify a subset whose joint capabilities cover the workflow and whose potential aggregate reward exceeds total execution and coordination cost. This procedure is formalized in Algorithm~\ref{alg:workflow-coalition} and yields a coalition that is both functionally sufficient and economically implementable.

Operationally, MCP remains unchanged. Once a feasible coalition is identified, the leader agent invokes each selected tool using standard MCP calls according to the derived workflow assignment. All schema validation, safety checks, and context handling proceed exactly as in existing MCP deployments. The novelty of \(\mathrm{C^{+}MCP}\) lies entirely in the pre-execution coordination layer, which determines coalition composition, tool participation, and effort allocation before any tool is invoked.

\section{Conclusion}
This paper advances scalable agentic intelligence by re-framing coordination among LLM-based agents as a networked systems problem. Rather than relying on centralized orchestration, we introduced the Internet of Agentic AI, a framework in which autonomous, heterogeneous agents distributed across cloud and edge infrastructure dynamically form coalitions to execute task-specific workflows.
By proposing incentive-compatible workflow-coalition feasibility framework and decentralized coordination algorithms, we demonstrated how scalable agentic AI can emerge from principled interaction among specialized agents operating under partial information and network constraints.

A central intuition of this work is that scalability in agentic AI is challenging to achieve by increasing model size. Instead, it requires architectural and economic mechanisms that enable agents to discover one another, coordinate effectively, and share rewards in an incentive-compatible manner. Our results show that coalition formation, workflow execution, and pricing (rewards) are intrinsically coupled problems, and need to be solved jointly to ensure operational feasibility and economic viability in open, distributed environments. The proposed framework provides a foundation for addressing these challenges while remaining compatible with existing agentic infrastructure.

The Internet of Agentic AI opens a broad set of future research directions. Immediate challenges include the design of pricing and incentive mechanisms \cite{yang2025pact}, market-based coordination protocols, responsibility attribution \cite{ge2024attributing}, and service-level agreements that govern interactions among independently operated agents. Integrating trust, reputation, and compliance into coalition formation also remains an open problem, particularly in domains such as healthcare and cybersecurity \cite{li2025texts}. 

%%%%%
\bibliography{reference}

@book{chalkiadakis2011computational,
  title={Computational Aspects of Cooperative Game Theory},
  author={Chalkiadakis, Georgios and Elkind, Edith and Wooldridge, Michael},
  year={2011},
  publisher={Morgan and Claypool Publishers}
}

@article{du2023multiagent,
  title={Multi-agent Debate with Large Language Models},
  author={Du, Yilun and Gao, Jianfei and Ghosh, Srijan and others},
  journal={arXiv preprint arXiv:2305.14325},
  year={2023}
}

@article{ghosh2016incentivizing,
  title={Incentivizing Participation in Peer-to-Peer Networks via Mechanism Design},
  author={Ghosh, Arpita and Hummel, Patrick},
  journal={Management Science},
  volume={62},
  number={7},
  pages={2085--2102},
  year={2016}
}

@article{liu2023camel,
  title={CAMEL: Communicative Agents for Mind Exploration of Large Scale Language Model Society},
  author={Liu, Zheng and Shen, Yujia and Li, Xiang and others},
  journal={arXiv preprint arXiv:2305.15002},
  year={2023}
}

@article{ma2024agentverse,
  title={AgentVerse: Facilitating Multi-Agent Collaboration and Exploring Emergent Behaviors},
  author={Ma, Yutian and Wang, Yujia and Lin, Tianxing and others},
  journal={arXiv preprint arXiv:2402.09629},
  year={2024}
}

@article{michalak2013efficient,
  title={Efficient Computation of the Shapley Value for Game-Theoretic Network Centrality},
  author={Michalak, Tomasz P. and Rahwan, Talal and Szczepanski, Pawe{\l} and Jennings, Nicholas R.},
  journal={Journal of Artificial Intelligence Research},
  volume={46},
  pages={607--650},
  year={2013}
}

@misc{openai2024agentic,
  title={Agentic AI: Enabling LLMs to Act Autonomously},
  author={OpenAI},
  note={https://openai.com/research/agentic-ai},
  year={2024}
}

@article{rahwan2007anytime,
  title={An Anytime Algorithm for Optimal Coalition Structure Generation},
  author={Rahwan, Talal and Ramchurn, Sarvapali D. and Jennings, Nicholas R.},
  journal={Journal of Artificial Intelligence Research},
  volume={34},
  pages={521--567},
  year={2009}
}

@article{saad2009coalitional,
  title={Coalitional Game Theory for Communication Networks},
  author={Saad, Walid and Han, Zhu and Debbah, Merouane and Hjorungnes, Are and Ba\c{s}ar, Tamer},
  journal={IEEE Signal Processing Magazine},
  volume={26},
  number={5},
  pages={77--97},
  year={2009}
}

@article{sandholm1999coalition,
  title={Coalition Structure Generation with Worst Case Guarantees},
  author={Sandholm, Tuomas and Larson, Kate and Andersson, Martin and Shehory, Onn and Tohmé, Fernando},
  journal={Artificial Intelligence},
  volume={111},
  number={1--2},
  pages={209--238},
  year={1999}
}

@article{shapley1953value,
  title={A Value for n-Person Games},
  author={Shapley, Lloyd S.},
  booktitle={Contributions to the Theory of Games},
  pages={307--317},
  year={1953},
  publisher={Princeton University Press}
}

@article{zhan2020learning,
  title={Learning Fair Allocation Rules in Cooperative Games},
  author={Zhan, Zeyuan and Bu, Fei and Yang, Bo and others},
  journal={arXiv preprint arXiv:2002.06142},
  year={2020}
}

@article{acharya2025agentic,
  title={Agentic AI: Autonomous Intelligence for Complex Goals--A Comprehensive Survey},
  author={Acharya, Deepak Bhaskar and Kuppan, Karthigeyan and Divya, B},
  journal={IEEE Access},
  year={2025},
  publisher={IEEE}
}

@inproceedings{wu2024autogen,
  title={Autogen: Enabling next-gen LLM applications via multi-agent conversations},
  author={Wu, Qingyun and Bansal, Gagan and Zhang, Jieyu and Wu, Yiran and Li, Beibin and Zhu, Erkang and Jiang, Li and Zhang, Xiaoyun and Zhang, Shaokun and Liu, Jiale and others},
  booktitle={First Conference on Language Modeling},
  year={2024}
}

@article{shu2024towards,
  title={Towards effective genAI multi-agent collaboration: Design and evaluation for enterprise applications},
  author={Shu, Raphael and Das, Nilaksh and Yuan, Michelle and Sunkara, Monica and Zhang, Yi},
  journal={arXiv preprint arXiv:2412.05449},
  year={2024}
}

@article{tran2025multi,
  title={Multi-agent collaboration mechanisms: A survey of llms},
  author={Tran, Khanh-Tung and Dao, Dung and Nguyen, Minh-Duong and Pham, Quoc-Viet and O'Sullivan, Barry and Nguyen, Hoang D},
  journal={arXiv preprint arXiv:2501.06322},
  year={2025}
}

@article{tian2025beyond,
  title={Beyond the Strongest LLM: Multi-Turn Multi-Agent Orchestration vs. Single LLMs on Benchmarks},
  author={Tian, Aaron Xuxiang and Zhang, Ruofan and Tang, Jiayao and Cho, Young Min and Li, Xueqian and Yi, Qiang and Wang, Ji and Zhang, Zhunping and Qi, Danrui and Li, Zekun and others},
  journal={arXiv preprint arXiv:2509.23537},
  year={2025}
}

@article{milne2020effectiveness,
  title={The effectiveness of artificial intelligence conversational agents in health care: systematic review},
  author={Milne-Ives, Madison and De Cock, Caroline and Lim, Ernest and Shehadeh, Melissa Harper and De Pennington, Nick and Mole, Guy and Normando, Eduardo and Meinert, Edward and others},
  journal={Journal of medical Internet research},
  volume={22},
  number={10},
  pages={e20346},
  year={2020},
  publisher={JMIR Publications Inc., Toronto, Canada}
}

@book{branzei2008models,
  title={Models in cooperative game theory},
  author={Branzei, Rodica and Dimitrov, Dinko and Tijs, Stef},
  year={2008},
  publisher={Springer}
}

@article{sarkar2022survey,
  title={A survey on applications of coalition formation in multi-agent systems},
  author={Sarkar, Samriddhi and Curado Malta, Mariana and Dutta, Animesh},
  journal={Concurrency and Computation: Practice and Experience},
  volume={34},
  number={11},
  pages={e6876},
  year={2022},
  publisher={Wiley Online Library}
}

@article{tu2022incentive,
  title={Incentive mechanisms for federated learning: From economic and game theoretic perspective},
  author={Tu, Xuezhen and Zhu, Kun and Luong, Nguyen Cong and Niyato, Dusit and Zhang, Yang and Li, Juan},
  journal={IEEE transactions on cognitive communications and networking},
  volume={8},
  number={3},
  pages={1566--1593},
  year={2022},
  publisher={IEEE}
}

@article{zhan2021survey,
  title={A survey of incentive mechanism design for federated learning},
  author={Zhan, Yufeng and Zhang, Jie and Hong, Zicong and Wu, Leijie and Li, Peng and Guo, Song},
  journal={IEEE Transactions on Emerging Topics in Computing},
  volume={10},
  number={2},
  pages={1035--1044},
  year={2021},
  publisher={IEEE}
}

@article{qi2025towards,
  title={Towards transparent and incentive-compatible collaboration in decentralized llm multi-agent systems: A blockchain-driven approach},
  author={Qi, Minfeng and Zhu, Tianqing and Zhang, Lefeng and Li, Ningran and Zhou, Wanlei},
  journal={arXiv preprint arXiv:2509.16736},
  year={2025}
}

@article{yang2025pact,
  title={PACT: A Contract-Theoretic Framework for Pricing Agentic AI Services Powered by Large Language Models},
  author={Yang, Ya-Ting and Zhu, Quanyan},
  journal={arXiv preprint arXiv:2505.21286},
  year={2025}
}

@article{li2025texts,
  title={From Texts to Shields: Convergence of Large Language Models and Cybersecurity},
  author={Li, Tao and Yang, Ya-Ting and Pan, Yunian and Zhu, Quanyan},
  journal={arXiv preprint arXiv:2505.00841},
  year={2025}
}

@article{ge2024attributing,
  title={Attributing responsibility in ai-induced incidents: A computational reflective equilibrium framework for accountability},
  author={Ge, Yunfei and Yang, Ya-Ting and Zhu, Quanyan},
  journal={arXiv preprint arXiv:2404.16957},
  year={2024}
}
\bibliographystyle{IEEEtran}

\end{document}